\documentclass{elsart}
\newcommand{\VEC}[1]{\textbf{\emph{#1}}}
\newcommand{\ud}{\mathrm{d}}
\begin{document}

\begin{frontmatter}

\title{Radiation by a relativistic charged particle in self-wakefield in periodic structure}
\author{Anatoliy Opanasenko\corauthref{cor1}}
\address{NSC Kharkov Institute of Physics and
        Technology, Akademicheskaya Str. 1, Kharkov, 61108, Ukraine}
\ead{Opanasenko@kipt.kharkov.ua} \corauth[cor1]{Tel. (380 0572) 35
62 96, fax (380 0572) 35 37 31}

\begin{abstract}
A new elementary mechanism of radiation  due to the oscillatory
character of a radiation reaction force appearing when a
relativistic charged particle moves along a periodic structure
without external fields is investigated. It is shown that the
non-synchronous spatial harmonics of Cherenkov-type radiation (CR)
can give rise to the oscillation of a particle which consequently
generates  undulator-type radiation (UR). In the spectral region,
where the diffraction of generated waves is essential, the
radiation manifests itself in the coherent interference of CR and
UR. A pure undulator-type radiation takes place only in the
wavelength range where the wave diffraction can be neglected. In
the case of coherent UR emitted by a bunch of $ N $ electrons, the
UR power is proportional to $ N^{4} $.
\end{abstract}

\end{frontmatter}

\section{Introduction}

In this paper we offer a new mechanism of radiation due to the
oscillatory character of a reaction force arising as a
relativistic charged particle moves through a periodic structure.
The impossibility of oscillatory motion of a free charged particle
in the self-field has been proved in~\cite{esF50}. However, it is
also well known  that a charged particle moving at a constant
velocity along the periodic structure emits the Cherenkov-type
radiation (or the diffraction radiation) \cite{AFL55}. The fields
of this radiation, called as wakefields, can be expressed as a
spatial-harmonics series expansion according to Floquet's theorem.
The action of synchronous spatial harmonics of the self-wakefields
on the particle results in energy losses associated with the
Cherenkov-type radiation. Under certain conditions, the
non-synchronous spatial harmonics give rise to the oscillatory
motion of the particle that consequently generates the
undulator-type radiation. This radiation is the subject of
discusson in this article.

\section{Methodology}

As a periodic structure, we will consider the vacuum corrugated
waveguide with a metallic surface. Such structures are commonly
used in rf linacs. Let a particle having the ultrarelativistic
velocity $\VEC{v}$, the charge $ e $ and the mass $ m $ moves
along the structure with the period $ D $. The longitudinal
component of the velocity $ v_{z} $, parallel to the structure
axis, is very close to the velocity of light $c$. The radiation
reaction force and the radiation power have to be found.

For calculating the radiation reaction force we will use the
approach developed in~\cite{GE59}. At first, we will suppose that
the charged particle is not a point charge, but is distributed
with density $\rho~=~ef[\VEC{r}-\VEC{r}(t)] $. Here $ \VEC{r}(t)$
is the radius vector of the center of mass of the particle and $
\int f(\VEC{r})\,\ud^{3}\VEC{r}=1 $. The equation of motion for
the center of mass is written as
{\setlength\arraycolsep{0pt}
\begin{eqnarray}\label{e:MOVE1}
\frac{\ud (m\gamma\VEC{v})}{\ud t}=e\int
  \left[
  \VEC{E}(\VEC{r},t)+ \frac{\VEC{v} \times
  \VEC{H}(\VEC{r},t)}{c}
  \right]
 f[\VEC{r}-\VEC{r}(t)]\,\ud^{3}\VEC{r},
\end{eqnarray}
\noindent where $  \gamma= 1\left/
\sqrt{1-\frac{v^{2}}{c^{2}}}\right.$ is the Lorentz factor, $
\VEC{E}(\VEC{r},t) $ and $\VEC{H}(\VEC{r},t) $ are, respectively,
the electrical and magnetic self-fields of the charge moving in
the periodic structure. These fields can be expressed in terms of
vector potential $\VEC{A}(\VEC{r},t)$  and scalar potential $
\Phi(\VEC{r},t) $ in the Coulomb calibration $ div\VEC{A}=0 $
\begin{equation}\label{e:FIELD}
  \VEC{E}=-\frac{1}{c}\frac{\partial \VEC{A}}{\partial t}-
  \nabla\Phi, \qquad \VEC{H}=rot\VEC{A}.
\end{equation}
The potential electric self-field $-\nabla\Phi$ does not influence
the motion of the center of mass for the distributed
charge~\cite{GE59}. So, it will suffice to find
$\textbf{A}(\textbf{r},t)$ which satisfies the wave equation
\begin{equation}\label{e:WAVE}
  \triangle{\VEC{A}}
   -\frac{1}{c^{2}}\frac{\partial^{2}\VEC{A}}{\partial t^{2}}
   =-\frac{4\pi}{c}\rho \VEC{v}
  +\frac{1}{c}\frac{\partial (\nabla\Phi)}{ \partial t}.
\end{equation}
We will seek for the solution of Eq.~(\ref{e:WAVE}) in the form of
the Fourier series
\begin{equation}\label{e:SOLFUR}
  \VEC{A}(\VEC{r}) =
     \mathrm{Re}\left[ \sum_{\lambda}
       q_{\lambda}(t)\VEC{A}_{\lambda}(\VEC{r})\right],
   \end{equation}
\noindent where $ q_{\lambda}(t) $ are certain unknown functions
of time $t$, $ \VEC{A}_{\lambda}(\VEC{r}) $ is a set of orthogonal
eigenfunctions of homogeneous Eq.~(\ref{e:WAVE}). \noindent
Inserting Eq.~(\ref{e:SOLFUR}) into Eq.~(\ref{e:WAVE}) yields the
equations for $ q_{\lambda}(t)$
\begin{eqnarray}\label{e:AMPL}
    \frac{\ud^{2}q_{\lambda}}{\ud t^{2}}
     +\omega_{\lambda}^{2}q_{\lambda}
    = \frac{e\VEC{v}(t)}{cV_{tot}}\int_{V_{tot}}
    \VEC{A}_{\lambda}^{\ast}(\VEC{r})f[\VEC{r}-\VEC{r}(t)]\,\ud^{3}\VEC{r},
\end{eqnarray}
\noindent where $ \omega_{\lambda} $ is a set of eigenfrequencies.
We solved these equations with the following initial conditions: $
q_{\lambda}(0)=0 $, $ \frac{dq_{\lambda}(t)}{dt}|_{t=0}=0 $. $
V_{tot}=MV_{cell} $ is the volume of the periodic structure. In
order to deal with a discrete set of waves we assume that the
structure is enclosed in a "periodicity box" containing
$M\to\infty$  cells of volume $V_{cell}$ \cite{AFL55}. Solving
Eq.~(\ref{e:AMPL}) and using the definitions (\ref{e:FIELD}), we
rewrite Eq.~(\ref{e:MOVE1}) as
\begin{eqnarray}\label{e:MOVE2}
&&\frac{\ud \left( m \gamma \VEC{v} \right)}{\ud t}
  =\VEC{F}(\VEC{v}(t),\VEC{r}(t), t )
   = - \frac{e^{2}}{4c^2 V_{tot} }\\
&&\times\!\!\sum\limits_\lambda ^{\omega _\lambda < c/r_0}
\left\{\!\left[
 \VEC{A}_\lambda (\VEC{r}( t))-\frac{\VEC{v}( t)\times rot\VEC{A}_\lambda (\VEC{r}( t))}{i\omega _\lambda }
\right] \right.
e^{i\omega_{\lambda}t}\int_{0}^{t}e^{-i\omega_{\lambda}t'}\VEC{v}(t')
     \VEC{A}_{\lambda}^{\ast}(\VEC{r})\,dt'
\nonumber\\
&&{}+\left[
 \VEC{A}_\lambda (\VEC{r}( t))+\frac{\VEC{v}( t)\times rot\VEC{A}_\lambda (\VEC{r}( t))}{i\omega _\lambda }
\right]
e^{-i\omega_{\lambda}t}\left.
\int_{0}^{t}e^{i\omega_{\lambda}t'}\VEC{v}(t')
     \VEC{A}_{\lambda}^{\ast}(\VEC{r})\,dt'
\right\}+ c.c.\nonumber
\end{eqnarray}
\noindent As the radiation reaction force  $\VEC{F}(t)$, unlike
the electromagnetic mass, does not depend on the particle size $
r_{0} $ (where $ r_{0} $ is meant in the laboratory frame of
reference), the distribution function $ f[\VEC{r}-\VEC{r}(t)] $
can be replaced by the Dirac $\delta$-function, and in the sum of
Eq.~(\ref{e:MOVE2}) only the frequencies  $ \omega _\lambda <
c/r_{0} $  are taken account  ~\cite{GE59}. The eigenfunctions of
the vector potential for the periodic structures with perfectly
conducting walls are usually given in the Floquet
form~\cite{AFL55}
\begin{equation}\label{e:EGENVEC}
  \VEC{A}_{\lambda}(\VEC{r})=
   \sum_{n=-\infty}^{\infty}
   \VEC{g}_{\lambda}^{(n)}(\VEC{r}_{\bot})e^{ih_{n}z}
   \end{equation}
\noindent where $\VEC{A}_{\lambda}(\VEC{r})$  is the periodic
function of $z$ with the period $D$,
$\VEC{g}_{\lambda}^{(n)}(\VEC{r}_{\bot})$ is the amplitude of the
$n$th spatial harmonic dependent on the transverse co-ordinates
$\VEC{r}_{\bot}$, $h$ is a discrete parameter multiple of $2\pi
/(MD)$ in the interval $ \left (-\pi /D \div \pi /D\right) $,
 $h_{n}=h+2\pi n/D $
 is the propagation
constant of the $n$th spatial harmonic.

The set of eigenfunctions~(\ref{e:EGENVEC}) for infinitely long
periodic waveguide is physically limited in frequency by the value
of electron plasma frequency $\omega_{e}$ in the metal. As is
known, if $\omega_{\lambda}\sim \omega_{e}$ , the conduction of
metal walls strongly falls off, and the diffraction conditions in
the periodic structure are disrupted. So, in the spectral region
$\omega_{e}~<~\omega_{\lambda}$, where the wave diffraction can be
neglected, the periodic waveguide can be considered as a free
space. In this part of frequency spectrum, the vector potential is
sought as expansion in terms of the plane waves
\begin{equation}\label{e:PLANEWAVE}
  \VEC{A}_{\lambda,l}(\VEC{r})=
   c\sqrt{4\pi} \VEC{a}_{\lambda l}
   e^{i\VEC{k}_{\lambda}\VEC{r}},
\end{equation}
\noindent where $\VEC{k}_{\lambda}$ is the wave propagation
vector; $\VEC{a}_{\lambda l}$ are the real unit vectors of
polarization $(l=1,2)$, perpendicular to $\VEC{k}_{\lambda}$.

\section{The zeroth-order wake force}

In the ultrarelativistic limit, the equation of
motion~(\ref{e:MOVE2}) can be solved by the method of successive
approximations. We will find non-relativistic corrections for the
particle velocity  $ v_{0}\approx c $ . As a zeroth order
approximation, we consider the uniform motion of the charged
particle parallel to the waveguide axis:
\begin{equation}\label{e:VRZERO}
  \VEC{v}= \VEC{v}_{0}=v_{0}\VEC{e}_{z},
   \qquad
  \VEC{r}(t)=\VEC{r}_{0 \bot}+\VEC{v}_{0}t
\end{equation}
Inserting Eqs.~(\ref{e:EGENVEC}) and (\ref{e:VRZERO}) (for the
frequency region $\omega_{\lambda}\ll\omega_{e}$) into the right
part of Eq.~(\ref{e:MOVE2}) we obtain the radiation reaction force
in the zeroth-order approximation in the form of a wake force
\begin{eqnarray}\label{e:FORCE1}
 \VEC{F}(t)= - e^{2}\sum_{p=-\infty}^{\infty}
    \VEC{w}^{(p)}e^{ip\Omega t} + c.c.
 \end{eqnarray}
\noindent where $\left.\Omega\equiv2\pi v_{0}\right/ D $ and the
amplitudes of spatial harmonics of the wake function are defined
as
\begin{eqnarray}\label{e:WAKEFUN}
&& \VEC{w}^{(p)}\equiv \frac{v_{0}D}{4c^2 V_{tot}}
\sum_{n=-\infty}^{\infty}\sum_{\lambda_{j}}
    \frac{g_{z \lambda_{j}}^{(n)*}}{\left|v_{0}-\frac{\ud \omega_{\lambda}}{\ud h}
    \right|_{\lambda=\lambda_{j}}}
    \nonumber \\
      &&\qquad\qquad\qquad\qquad\qquad
       \times
      \left[
       \VEC{g}_{z \lambda_{j}}^{(n+p)}
       -i\frac{v_{0}}{\omega_{\lambda_{j}}}\nabla_{\bot}g_{z \lambda_{j}}^{(n+p)}
       -\frac{\Omega p}{\omega_{\lambda_{j}}}\VEC{g}_{\bot \lambda_{j}}^{(n+p)}
       \right]
\end{eqnarray}
\noindent Hereinafter, the amplitudes of spatial harmonics are
taken at $\VEC{r}=\VEC{r}_{0 \bot}$ as
 $\VEC{g}_{\lambda}^{(n)}~\equiv~\VEC{g}_{\lambda}^{(n)}(\VEC{r}_{0
\bot})$, $ \omega_{\lambda_{j}} $ satisfies the resonance
conditions $ hv_{0}~-~\omega_{\lambda} = n \Omega $. The wake
force~(\ref{e:FORCE1}) is the periodic function of time with the
period $ \left. D \right/ v_{0} $. The synchronous harmonic of the
force $ -2e^{2}w_{z}^{(0)} $ defines the energy losses associated
with Cherenkov-type radiation. The power of this radiation
$2v_{0}e^{2}w_{z}^{(0)}$ agrees with the one given in
Ref.~\cite{AFL55}. As it is easily seen, the transverse component
of the synchronous harmonic of the wake force equals zero, as
$\VEC{w}_{\bot}^{(0)}=0 $.

In the range of $ \omega_{e}~<~\omega_{\lambda} $, where the
structure is supposed as a free space, there is no radiation in
the zeroth order approximation (at $ \VEC{v}_{0}=const $).

\section{The first-order approximation}

    If the charged particle moves off-axis, it experiences the action of
the transverse component of nonsynchronous harmonics of the wake
force ($ \VEC{w}_{\bot}^{(p)} \neq 0 $). So, we will find
non-relativistic corrections for both the velocity $ \VEC{v}_{0} $
and the radius vector $\VEC{r}_{\bot}$ of the off-axis particle
that are caused by the periodic transverse wake force. We assume
that the change in the longitudinal velocity is negligible.
Putting Eq.~(\ref{e:FORCE1}) into the equation of
motion~(\ref{e:MOVE2}) we correct the law of motion
\begin{eqnarray}\label{e:VELOS}
 &&\VEC{v}(t)=\VEC{v}_{0}+\VEC{v}_{\bot}(t)=\VEC{v}_{0}+
 ic\sum_{p\neq 0}^{\infty}\frac{\VEC{b}^{(p)}}{p}  e^{ip\Omega t},\\
 &&\VEC{r}(t)=\VEC{r}_{0 \bot}\!+\VEC{v}_{0}t+\delta \VEC{r}_{\bot}(t)=\VEC{r}_{0
 \bot}\!+\VEC{v}_{0}t+
 \frac{c}{\Omega}\sum_{p \neq 0}^{\infty}\frac{\VEC{b}^{(p)}}{p^{2}}e^{ip \Omega
 t},\label{e:RADIUS}
\end{eqnarray}
\noindent where $ \VEC{b}^{(p)} $  is the dimensionless vector
\begin{equation}\label{e:b}
  \VEC{b}^{(p)}\equiv \frac{2e^{2}}{mc\gamma\Omega}
     \left ( \VEC{w}_{\bot}^{(p)}+\VEC{w}_{\bot}^{(-p)*}\right).
\end{equation}
\noindent The absolute value $ \mid \VEC{b}^{(p)}\mid $ is the
small parameter.

Substituting Eqs.~(\ref{e:VELOS}),  (\ref{e:RADIUS}) and
(\ref{e:EGENVEC}) into Eq.~(\ref{e:MOVE2}), and multiplying it by
$ \VEC{v}$, we obtain the power radiation within the accuracy
  $\mid \VEC{b}^{(p)}\mid^{2} $ in the range  $\omega_{\lambda}\ll\omega_{e}$
\begin{eqnarray}\label{e:POWER1}
&&P\equiv-\lim_{t\to \infty}\frac{1}{t}
    \int_{0}^{t}\VEC{v}( t)\VEC{F}\left( \VEC{v}( t) ,\VEC{r}(t), t\right)\,\ud t
     \nonumber\\
    &&\qquad\quad=\frac{e^{2}\pi v_{0}}{2cV_{tot} }\sum\limits_\lambda
    \sum_{n=-\infty}^{\infty}
\left[\vphantom{\sum}\delta(h_{n}v_{0}-\omega_{\lambda})\!
+\delta(h_{n}v_{0}+\omega_{\lambda}) \vphantom{\sum}\right ]\nonumber \\
&&\qquad\qquad\qquad\qquad\times
\left|g_{z\lambda}^{(n)}+\sum_{p\neq0}\frac{\VEC{b}^{(p)}}{2p}
\left(\frac{c}{p\Omega}\nabla_{\bot}g_{z \lambda}^{(n+p)}
-\frac{ic}{v_{0}}\VEC{g}_{\bot \lambda}^{(n+p)}\right)\right|^{2}.
\end{eqnarray}
Replacing the summation over discrete $h$ by integration at
$M\to\infty$ in Eq.~(\ref{e:POWER1}) we find that
\begin{eqnarray}\label{e:POWER2}
&&P=\frac{e^{2}v_{0}D}{2cV_{cell}}
\sum_{n=0}^{\infty}\sum_{\lambda_{j}}
 \frac{1}{\left|v_{0}-\frac{\ud \omega_{\lambda}}{\ud h}
 \right|_{\lambda=\lambda_{j}}} \nonumber \\
&&\qquad\qquad\qquad\qquad\times \left|
g_{z\lambda_{j}}^{(n)}\!+\!\sum_{p\neq0}\frac{\VEC{b}^{(p)}}{2p}
\!\left(\frac{c}{p\Omega}\nabla_{\bot}g_{z \lambda_{j}}^{(n+p)}
-\frac{ic}{v_{0}}\VEC{g}_{\bot \lambda_{j}}^{(n+p)}\right)
\right|^{2},
\end{eqnarray}
\noindent where $ \omega_{\lambda_{j}} $ satisfies the resonance
conditions $ hv_{0}~-~\omega_{\lambda} = n \Omega $ .

Eq.~(\ref{e:POWER2}) shows that in the region
$\omega_{\lambda}\ll\omega_{e}$ there is coherent interference
between the Cherenkov-type radiation and the undulator-type
radiation, that is caused by the oscillation of the particle in
the nonsynchronous harmonics self-wakefield. As is evident from
Eqs.~(\ref{e:POWER2}) and (\ref{e:b})  in this frequency range the
total radiation power tends to the CR power with increasing
$\gamma$.

Let us next consider the radiation of the charge particle in the
range $\omega_{e}~<~\omega_{\lambda}$. In analogy with
Eq.~(\ref{e:POWER1}), substituting Eqs.~(\ref{e:VELOS}),
(\ref{e:RADIUS}) and (\ref{e:PLANEWAVE}) into Eq.~(\ref{e:MOVE2}),
we can obtain the power of the pure undulator-type radiation
\begin{eqnarray}\label{e:POWERSYN1}
&&P_{U}\equiv-\lim_{t\to \infty}\frac{1}{t}
    \int_{0}^{t}\VEC{v}( t)\VEC{F}
    \left( \VEC{v}( t) ,\VEC{r}(t), t\right)\,\ud t \nonumber\\
     &&\qquad=\frac{e^{2}c^{2}\pi^{2}}{2V_{tot} }\!\sum\limits_\lambda
    \!\sum_{p\neq 0}\!\sum_{l=1}^{2} \left| \frac{\VEC{a}_{\lambda l}
\VEC{b}^{(p)}}{p}- \frac{Da_{z \lambda l}\VEC{k}_{\bot
\lambda}\VEC{b}^{(p)}}{2\pi p^{2}}
\right|^{2} \nonumber\\
&&\qquad\qquad\qquad\qquad\times  \left[ \vphantom{\sum}
\delta\left(k_{z\lambda}v_{0}-p\Omega-\omega_{\lambda}\right)
+\delta\left(k_{z\lambda}v_{0}-p\Omega+\omega_{\lambda}\right)
 \vphantom{\sum}\right ].
\end{eqnarray}
\noindent Here to simplify the calculations, we consider the
oscillation of the particle in the dipole limit
\begin{eqnarray}\label{e:LIMIT}
\VEC{k}_{\lambda}\delta\VEC{r}_{\bot}(t)\ll 2\pi.
\end{eqnarray}

Considering Eq.~(\ref{e:LIMIT}) and the wave dispersion in a free
space $\left ( \omega_{\lambda}=ck_{\lambda}\right)$, we go from
the summation over $\lambda$ in Eq.~(\ref{e:POWERSYN1}) to
integration over $\omega$ at $V_{tot}\to\infty $
\begin{eqnarray}\label{e:POWERSYN2}
&&P_{U}=\frac{e^{2}}{16\pi c}
    \int_{0}^{2\pi}\ud \varphi
    \int_{0}^{\pi}\sin \theta \ud \theta
    \int_{\omega_{e}}^{\omega_{max}<c/r_0}\omega^{2}\ud \omega
    \nonumber \\
&&\times\sum_{p\neq 0}
 \left\{ \frac{\mid b_{x}^{(p)}\mid^{2}}{p^{2}}
\left[ 1-R(\omega,\theta,p)\sin^{2}\theta\cos^{2}\varphi \right]
 \right.\nonumber \\
&&\qquad{}+\frac{\mid b_{y}^{(p)} \mid^{2}}{p^{2}} \left[
1-R(\omega,\theta,p)\sin^{2}\theta\sin^{2}\varphi \right]
\nonumber \\
&&\qquad{}-\left. \mathrm{Re}\left[\frac{b_{x}^{(p)}b_{y}^{(p)}}
{p^{2}}\right]R(\omega,\theta,p)\sin^{2}(2\varphi)\sin^{2}\theta
\right\}\nonumber \\
&& \qquad\times \left\{\vphantom{\sum} \delta\left [
\omega(\beta_{0}\cos\theta-1)-p\Omega\right ] \right. \left.
 +\delta\left [
\omega(\beta_{0}\cos\theta+1)-p\Omega\right ]
 \vphantom{\sum}\right\},
\end{eqnarray}
\noindent where  $ R(\omega,\theta,p)\equiv\left(
1-\frac{\omega}{p\Omega}\beta_{0}\cos\theta
\right)^{2}-\left(\beta_{0}\frac{\omega}{p\Omega}\right)^{2} $,
 $\theta$ is the angle between the wave vector $\VEC{k}$ and
the $Oz$ axis, $\varphi$ is the angle between the $Îx$ axis  and
the $xOy$-plane projection of $\VEC{k}$, $ \beta_{0}=v_{0}/c$.

For $\omega_{e}~<~\omega_{\lambda}$ it is of interest to consider
the radiation of the high energy charged particle satisfying the
condition $\omega_{e}~\ll~\Omega\gamma^{2}$. In this case,
integrating over $\omega$, $\theta$, $\varphi$ in
Eq.~(\ref{e:POWERSYN2}) and replacing $\VEC{b}^{(p)}$ from
Eq.~(\ref{e:b})  we find the total power of pure undulator-type
radiation
\begin{eqnarray}\label{e:POWERSYN3}
P_{U}=\frac{4e^{6}}{3m^{2}c^{3}}\gamma^{2}\sum_{p=1}^{p\ll
p_{lim}}
\left|\VEC{w}_{\bot}^{(p)}+\VEC{w}_{\bot}^{(-p)*}\right|^{2},
\end{eqnarray}
\noindent where the number of harmonics in the sum is limited by
the condition (\ref{e:LIMIT}) resulting in $p\ll
p_{lim}=2\pi\gamma/ max|b^{(p)}|$. As it follows from
Eq.~(\ref{e:POWERSYN3}), the power grows as $\sim \gamma^{2}$, so
in the region $\omega_{e}~\ll~\omega_{\lambda}$ the UR power can
exceed the CR power emitted in the spectral region
$\omega_{\lambda}~\ll~\omega_{e}$.

It should also be stated that, if instead of the above considered
point particle there is a bunch of $N$  electrons with
longitudinal and transverse dimensions ($\sigma_{z}$ and
$\sigma_{\bot}$) which satisfy the both conditions  $\sigma_{z}\ll
D/(2q\gamma^{2})$ and $\sigma_{\bot}\ll D/(2q\gamma)$, then   the
radiation is coherent in the frequency region
$\omega~<~2q\Omega\gamma^{2}$ . Moreover, as it follows from
Eq.~(\ref{e:POWERSYN3}),  the UR power is proportional to $N^{4}$
in the range $\omega_{e}\ll\omega~<~2q\Omega\gamma^{2}$
\begin{eqnarray}\label{e:POWERSYNCOH}
P_{U}=\frac{4e^{6}N^{4}}{3m^{2}c^{3}}\gamma^{2}\sum_{p=1}^{q}
\left|\VEC{w}_{\bot}^{(p)}+\VEC{w}_{\bot}^{(-p)*}\right|^{2}.
\end{eqnarray}

\section{Conclusions}

The new radiation mechanism considered above may be of use in
undulators based on periodic structures without external fields,
where the non-synchronous wake-harmonics of an electron bunch act
as pump waves. These wakefield undulators require no magnetic
fields or rf sources needed in present-day FEL. Note also that the
undulator-type radiation power is proportional to $\gamma^{2}$.
So, in the future high energy electron rf linacs, in view of
deviation of a beam from the linacs axis, because of the coherent
betatron oscillation of the beam in a focussing system, the
interaction of electrons with the spatial non-synchronous
harmonics of both an accelerating mode ~\cite{anO00} and a
wakefield may result in the electron energy loss associated with
the spontaneous undulator-type radiation.

\section{Acknowledgements}

The author is grateful to Academician Ya.B.~Fainberg for the
proposed method of solution and for fruitful discussions.

\end{document}